\documentstyle[amssymb,pra,aps,epsfig,12pt]{revtex}

\newcommand{\be}{\begin{equation}}
\newcommand{\ee}{\end{equation}}
\newcommand{\br}{\begin{eqnarray}}
\newcommand{\er}{\end{eqnarray}}

\newcommand{\bd}{\begin{displaymath}}
\newcommand{\ed}{\end{displaymath}}

\newcommand{\bfig}{\begin{figure}}
\newcommand{\efig}{\end{figure}}

\def\3cdot{\cdot \cdot \cdot}

\def\om0{\omega _0}
\def\Om0{\Omega _0}

\def\text#1{{\rm{#1}}}

\def\->{\rightarrow}
\def\=>{\Rightarrow}
\def\-->{\longrightarrow}
\def\==>{\Longrightarrow}

\def\pr{^\prime}

\def\pr2{^{\prime\prime}}

\def\bfig{\begin{figure}}
\def\efig{\end{figure}}

\begin{document}
\title{{\Large High-fidelity teleportation of entanglements of running-wave field
states}}
\author{R. M. Serra$^{1}$\thanks{%
E-mail: serra@df.ufscar.br}, C. J. Villas-B\^{o}as$^{1}$, N. G. de Almeida$%
^{2}$, and M. H. Y. Moussa$^{1}$\thanks{%
E-mail: miled@df.ufscar.br}.}
\address{$^{1}$Departamento de F\'{\i}sica, Universidade Federal de S\~{a}o\\
Carlos, P.O. Box 676, S\~{a}o Carlos, 13565-905, S\~{a}o Paulo, Brazil. \\
$^{2}$Departamento de Matem\'{a}tica e F\'{\i}sica, Universidade\\
Cat\'{o}lica de Goi\'{a}s, P.O. Box 86, Goi\^{a}nia,74605-010, Goi\'{a}s,\\
Brazil.}
\maketitle

\begin{abstract}
We describe a scheme for the teleportation of entanglements of zero- and
one-photon running-wave field states. In addition to linear optical
elements, Kerr nonlinearity is also employed so as to achieve a $100\%$
probability of success in the ideal case. A comprehensive phenomenological
treatment of errors in the domain of running-wave physics, for linear and
nonlinear optical elements, is also given, making it possible to calculate
the fidelity of the teleportation process. A strategy for carrying out the
Bell-type measurement which is able to probe the absorption of photons in
the optical elements is adopted. Such strategy, combined with usually small
damping constants characterizing the optical devices, results in a high
fidelity for the teleportation process. The feasibility of the proposed
scheme relies on the fact that the Kerr nonlinearity it demands can be
achieved through the recently reported ultraslow light propagation in cold
atomic media [Phys. Rev. Lett. {\bf 84}, 1419 (2000); Phys. Rev. A {\bf 65},
033833 (2002)].

{\bf Journal Ref. }J. Opt. B: Quantum Semiclass. Opt. {\bf 4}, 316 (2002)

{\bf arXiv:} http://arXiv.org/abs/quant-ph/0204057
\end{abstract}

\section{Introduction}

The property of nonlocality exhibited by entangled states, first pointed out
by Einstein, Podolsky and Rosen (EPR) as the cornerstone for their argument
against the uncertainty principle \cite{EPR}, has since then largely been
invoked for investigating the foundations of quantum mechanics. The program
inaugurated by the confrontation of EPR with the standard Copenhagen
interpretation of quantum mechanics, in keeping with the reinterpretation by
Bohm \cite{Bohm} of the gedankenexperiment designed by EPR, came to a head
with the possibility of an empirical test of nonlocality formulated by Bell 
\cite{Bell}. In the end, about two decades were devoted to the quest for
experimental demonstration of nonlocality through the violation of Bell's
inequalities \cite{Aspect}. Despite the majority of the experiments having
confirmed nonlocality, experimental loopholes have been pointed out which
have to be circumvented for an impartial conclusion, which hopefully will be
arrived at \cite{Chiao}.

However, in the last decade a variety of potential applications of
nonlocality have been devised which have definitely moved the focus of the
nonlocality phenomenon. From its original purely theoretical role in the
foundations of quantum mechanics, the nonlocality phenomenon together with
other fundamental quantum processes seems to be about to inaugurate a novel
technology for quantum communication \cite{Communication} and computation 
\cite{Computation}. Basically, such a possibility relies on the discovery
made by Shor \cite{Shor} that quantum information processing, involving
two-state systems as quantum bits, provides a means of integer factorization
much more efficiently than conventional computation. (Here we stress the
recent remarkable experimental realization of Shor's quantum factoring
algorithm using nuclear magnetic resonance, reported by the Solid State and
Photonics Laboratory of IBM, California \cite{Chuang}.) The information thus
processed and transmitted in a quantum logical processor consists of
arbitrary superpositions of quantum states instead of classical bits. The
interference phenomena characteristic of quantum superposition states allows
parallel computation paths which can reinforce or cancel one another,
depending on their relative phase \cite{Bennett}. Besides being
indispensable for correlating the input qubits in a quantum gate, the role
of entanglement and nonlocality in a quantum processor provides a striking
difference from any classical operation: states can be transmitted from one
node of a network to another by quantum teleportation. Discovered by Bennett
et al. \cite{Teleportation}, teleportation is a process in which a
superposition state, a qubit, is teleported from one quantum system to
another, over arbitrary distances, via dual classical and EPR channels, and
it furnishes a critical ingredient for the implementation of a quantum
logical processor \cite{Bennett}. It is worth noting that the state to be
teleported is destroyed during the required Bell state measurement; owing to
the linearity of quantum mechanics, a quantum state cannot be cloned \cite
{Zurek}, preventing the nonlocality phenomenon from being employed for
superluminal communication. However, quantum states can be transferred from
one system to another and even interchanged between different quantum
systems \cite{TI}, despite of the impossibility of cloning.

Since its proposition, the teleportation phenomenon has attracted great
attention and a number of different protocols have been suggested for
practical implementation of the process in the cavity QED \cite{David},
trapped ions \cite{Solano} and running wave domains \cite{Kimble1,Celso}.
Teleportation of entanglements \cite{TI,Loock,Kim,Lee,vanEnk,Ikram,Martini}
and $N(>2)$-dimensional states \cite{Miled}, of major interest for
information processing, has also been addressed, in addition to the
originally proposed protocol for teleporting qubits \cite{Teleportation}.
The process has also been demonstrated in experiments through photon
polarized states \cite{Zeilinger,Boschi,Shih}, and also through a
``Schr\"{o}dinger cat''-like state generated by parametric down-conversion
as a running wave{\it \ }\cite{Kimble2}. The Bell state measurement,
performed on the Bell operator basis consisting of four states -- a $%
2\otimes 2$ dimensional basis for the particle whose state is to be
teleported plus one of the two particles composing the quantum channel \cite
{Teleportation} -- constitutes the main experimental challenge. In the
Innsbruck experiment \cite{Zeilinger}, designed to manipulate only the
polarization state of single photon pulses, only one of the four Bell states
is discriminated, resulting in a success rate not larger than $25\%$.
Employing the entanglement between the spacial and polarization degrees of
freedom of a photon, the Rome experiment \cite{Boschi} was able to
distinguish the four Bell states allowing, in the ideal case, a $100\%$
success rate for teleportation. The experimental implementation reported in 
\cite{Boschi} was based on an approach suggested in \cite{Popescu},
different from the original protocol in \cite{Teleportation}: the
polarization degree of freedom of one of the photons composing the
EPR-quantum channel, a {\bf k}-vector entangled state, was used to prepare
the unknown state to be teleported. Such a strategy avoids the difficulties
associated with having three photons, making the Bell measurement more
straightforward. By accomplishing the Bell state measurements on the basis
of nonlinear interactions, the Baltimore experiment \cite{Shih}, which
follows exactly the original protocol by Bennett {\it et al. }\cite
{Teleportation}, also achieves, in principle, a $100\%$ success rate for
teleporting a polarization state. In the Caltech experiment \cite{Kimble2},
the teleportation of a state of continuous quantum variables was achieved
with a fidelity of $0.58\pm 0.02$, demonstrating the nonclassical character
of the process: the critical fidelity of $0.5$ is the classical bound
attainable in the absence of quantum correlation.

Here we introduce a high-fidelity technique for teleportation of
entanglements of zero- and one-photon running-wave states. The teleportation
apparatus is based on Mach-Zehnder (MZ) interferometry with a
phase-sensitive element, a cross-Kerr medium, allowing a $100\%$ probability
of success in the ideal case. A scheme to teleport similar entanglements of
running-wave states, employing linear optical elements at the expense of a $%
50\%$ probability of success, was recently suggested \cite{Kim,Lee}. In this
connection we note that the feasibility of our proposed scheme relies on a
recent demonstration of ultraslow light propagation in cold atomic media 
\cite{Hau}, which opened the way for the realization of significant
conditional phase shifts between two traveling single photon pulses.\cite
{Lukin,Petrosyan}

The teleportation and decoherence of entangled coherent running-wave states
has also been addressed \cite{vanEnk}, the probability of success in this
case being $50\%$. In the cavity QED domain, protocols have been reported
that teleport two-particle entangled atomic states \cite{TI}, multiparticle
atomic states, and entangled field states inside high-$Q$ cavities \cite
{Ikram}. We stress that experimental teleportation of an entangled qubit has
recently been achieved \cite{Martini}. Employing linear optics, the
experimental accomplishment in \cite{Martini}, following lines suggested in 
\cite{Kim}, had a $25\%$ probability of success. Efforts towards the
achievement of $50\%$ success rate (as described in \cite{Kim}) are in
progress \cite{Martini}.

Our experimental setup to teleport an entangled state consists basically of
three stations, sketched in Fig. 1(a), employing $50/50$ symmetric beam
splitters, Kerr media and photodetectors. In addition to the entanglement to
be teleported, which we assume to be already prepared when injected through
channels $1$ and $2$ ($|\Omega \rangle _{12}$), the quantum channel is
engineered by the apparatus sketched in Fig. 1(b). This{\it \ quantum
channel station} consists of a beam splitter ($BS_1$), a MZ interferometer,
composed of a pair of beam splitters ($BS_2$, $BS_3$), and a Kerr medium ($%
KM_1$). The latter, used to entangle the output modes $4$ and $5$ of $BS_1$
and $BS_2$, respectively, is an ingredient crucial to the generation of the
quantum channel (an entanglement of modes $3$,$4$,$5$, and $6$) required for
our proposed teleportation protocol. A part from the photodetectors, {\it %
Alice's station}, shown in Fig. 1(c), consists of the same ingredients used
to engineer the quantum channel. A MZ interferometer ($BS_4$, $BS_5$) is
disposed so as to receive modes $3$ and $4$ of the quantum channel, which is
coupled to the entanglement to be teleported through a Kerr medium ($KM_2$).
Finally, $BS_6$ is employed to prepare the whole entanglement (involving the
quantum channel and the state to be teleported) for the Bell-type
measurement carried out with photodetectors $D_i$ ($i=1$,$2$,$3$,$4$). In 
{\it Bob's station}, illustrated in Fig. 1(d), another MZ interferometer ($%
BS_7$, $BS_8$) and three phase plates are disposed so as to accomplish all
the rotations required to convert the teleported state into a replica of the
original entangled state $|\Omega \rangle _{12}$, allowing a $100\%$
probability of success in the ideal case.

It should be stressed that the fidelity of the scheme proposed for the
teleportation of an entangled state is estimated by taking into account the
noise introduced by dissipation in both the optical systems, beam splitters
and Kerr media. We present here a phenomenological approach that allows for
the influence of damping in Kerr media, reasoning by analogy with the
treatment of lossy beam splitters given in \cite{Jeffers}. The efficiency of
the photodetectors has also been included using the relations previously
established in \cite{Celso}. We have pursued a strategy of carrying out the
Bell measurement in such a way as to probe the absorption of photons in the
optical elements placed between channels $1$ to $4$. This strategy, combined
with the usually small damping constants characterizing beam splitters and
Kerr media and the high efficiency of the photodetectors, results in a
high-fidelity teleportation process.

\section{Teleportation of Entangled States: Ideal Process}

Let us first introduce the situation where losses are disregarded and
describe the physical operations taking place in the optical elements. The
general relationships between the input and output operators $\widehat{%
\alpha }$,$\widehat{\beta }$ (see Fig. 2(a)), arising from the unitary
operator $\widehat{U}_{BS}=\exp \left[ i\theta \left( \widehat{\alpha }%
_{in}^{\dagger }\widehat{\beta }_{in}+\widehat{\beta }_{in}^{\dagger }%
\widehat{\alpha }_{in}\right) \right] $ describing the action of an ideal
symmetric $BS$, are written as

\begin{mathletters}
\begin{eqnarray}
\widehat{\alpha }_{out} &=&t\widehat{\alpha }_{in}+r\widehat{\beta }_{in}%
{\rm {,}}  \label{eq1a} \\
\widehat{\beta }_{out} &=&t\widehat{\beta }_{in}+r\widehat{\alpha }_{in}{\rm 
{,}}  \label{eq1b}
\end{eqnarray}
where $t=\cos (\theta )$ and $r=i\sin (\theta )$ are the beam-splitter
transmission and reflection coefficients satisfying $\left| t\right|
^2+\left| r\right| ^2=1$. For a 50/50 symmetric beam splitter, where{\bf \ }$%
\theta =\pi /4$, $t=\left| r\right| =1/\sqrt{2}$. These coefficients, and
thus the operators, depend on the frequency of the fields and here a
monochromatic source is assumed.

The coupling between the input and output operators $\widehat{\alpha }$,$%
\widehat{{\Bbb \beta }}$ (see Fig. 2(b)), representing the fields crossing a
Kerr medium, follows from the action of the unitary operator $\widehat{U}%
_{Kerr}=\exp \left( -i\chi \tau \widehat{\alpha }_{in}^{\dagger }\widehat{%
\alpha }_{in}\widehat{\beta }_{in}^{\dagger }\widehat{\beta }_{in}\right) $,
and is given by 
\end{mathletters}
\begin{mathletters}
\begin{eqnarray}
\widehat{\alpha }_{out} &=&\exp \left( i\chi \tau \widehat{\beta }%
_{in}^{\dagger }\widehat{\beta }_{in}\right) \widehat{\alpha }_{in}{\rm {,}}
\label{eq2a} \\
\widehat{\beta }_{out} &=&\exp \left( i\chi \tau \widehat{\alpha }%
_{in}^{\dagger }\widehat{\alpha }_{in}\right) \widehat{\beta }_{in}{\rm {,}}
\label{eq2b}
\end{eqnarray}
The conditional phase shift $\phi =\chi \tau $ depends upon the third-order
nonlinear susceptibility $\chi $ for the optical Kerr effect and the
interaction time $\tau $ within the Kerr medium. Finally, phase plates are
used to introduce an adjustable phase shift in the output field states and
photodetectors are employed for the projective Bell measurement.

The entangled state to be teleported can easily be prepared from a
single-photon field incident on a $BS$ with arbitrary unknown transmission
and reflection coefficients ${\cal C}_1$ and ${\cal C}_2$, satisfying $%
\left| {\cal C}_1\right| ^2+\left| {\cal C}_2\right| ^2=1$. Such a state,
injected through the input modes $1$ and $2$, can be written 
\end{mathletters}
\begin{equation}
\left| \Omega \right\rangle _{12}={\cal C}_1\left| 0\right\rangle _1\left|
1\right\rangle _2+{\cal C}_2\left| 1\right\rangle _1\left| 0\right\rangle _2%
{\rm {.}}  \label{eq3}
\end{equation}

Simultaneously to the preparation of the entanglement to be teleported, the
quantum channel is prepared from single-photon fields $\left| 1\right\rangle
_{4}$ and $\left| 1\right\rangle _{5}$ incident on $BS_{1}$ and $BS_{2}$,
respectively. It is easily verified from Eqs. (\ref{eq1a}, \ref{eq1b}) and (%
\ref{eq2a}, \ref{eq2b}) that the entangled state resulting in the output
modes of the quantum channel station (Fig. 1(b)) is written, apart from an
irrelevant phase factor, as

\begin{equation}
\left| \Upsilon \right\rangle _{3456}=\frac{1}{\sqrt{2}}\left( \left|
1\right\rangle _{3}\left| 0\right\rangle _{4}\left| 0\right\rangle
_{5}\left| 1\right\rangle _{6}+\left| 0\right\rangle _{3}\left|
1\right\rangle _{4}\left| 1\right\rangle _{5}\left| 0\right\rangle
_{6}\right) {\rm {,}}  \label{eq4}
\end{equation}
where the interaction parameter in the Kerr medium has been adjusted so that 
$\chi \tau =\pi $ \cite{Lukin,Petrosyan}. It must be stressed that the Kerr
medium is indispensable for engineering the correlated channels in Eq. (\ref
{eq4}). The product of the entanglement to be teleported and the quantum
channel, $\left| \Omega \right\rangle _{12}\otimes \left| \Upsilon
\right\rangle _{3456}$, can be expanded as

\begin{eqnarray}
\left| \Xi \right\rangle _{123456} &=&\frac 12\left\{ \left| \Psi
^{+}\right\rangle _{1234}\left( {\cal C}_1\left| 0\right\rangle _5\left|
1\right\rangle _6+{\cal C}_2\left| 1\right\rangle _5\left| 0\right\rangle
_6\right) +\left| \Psi ^{-}\right\rangle _{1234}\left( {\cal C}_1\left|
0\right\rangle _5\left| 1\right\rangle _6-{\cal C}_2\left| 1\right\rangle
_5\left| 0\right\rangle _6\right) \right.  \nonumber \\
&&\left. +\left| \Phi ^{+}\right\rangle _{1234}\left( {\cal C}_2\left|
0\right\rangle _5\left| 1\right\rangle _6+{\cal C}_1\left| 1\right\rangle
_5\left| 0\right\rangle _6\right) -\left| \Phi ^{-}\right\rangle
_{1234}\left( {\cal C}_2\left| 0\right\rangle _5\left| 1\right\rangle _6-%
{\cal C}_1\left| 1\right\rangle _5\left| 0\right\rangle _6\right) \right\} 
{\rm {.}}  \label{eq5}
\end{eqnarray}
We have introduced the complete set of $4$-particle eigenstates of the Bell
operators $\widehat{{\cal O}}_k=\left| \Theta _k^{\pm }\right\rangle
_{1234}\left\langle \Theta _k^{\pm }\right| $ ($\sum_k\widehat{{\cal O}}_k=1$%
), defined by

\begin{equation}
\left| \Theta _k^{\pm }\right\rangle _{1234}=\frac 1{\sqrt{2}}\left( \left| 
{\rm {bin}}\left( k\right) \right\rangle _{1234}\pm \left| {\rm {bin}}\left(
15-k\right) \right\rangle _{1234}\right) ,{\rm {\quad }}k=0,1,...,15,
\label{eq6}
\end{equation}
where bin$\left( k\right) $ refers to the $4$-bit binary representation of
the integer $k$. For the present teleportation protocol we are concerned
only with four states out of the complete Bell basis $\left| \Theta _k^{\pm
}\right\rangle _{1234}$, those employed for the expansion in (\ref{eq5}):

\begin{mathletters}
\begin{eqnarray}
\left| \Psi ^{\pm }\right\rangle _{1234} &=&\left| \Theta _6^{\pm
}\right\rangle _{1234}=\frac 1{\sqrt{2}}\left( \left| 0\right\rangle
_1\left| 1\right\rangle _2\left| 1\right\rangle _3\left| 0\right\rangle
_4\pm \left| 1\right\rangle _1\left| 0\right\rangle _2\left| 0\right\rangle
_3\left| 1\right\rangle _4\right) ,  \label{eq7a} \\
\left| \Phi ^{\pm }\right\rangle _{1234} &=&\left| \Theta _5^{\pm
}\right\rangle _{1234}=\frac 1{\sqrt{2}}\left( \left| 0\right\rangle
_1\left| 1\right\rangle _2\left| 0\right\rangle _3\left| 1\right\rangle
_4\pm \left| 1\right\rangle _1\left| 0\right\rangle _2\left| 1\right\rangle
_3\left| 0\right\rangle _4\right) .  \label{eq7b}
\end{eqnarray}
Therefore, measurements on the output fields $1$,$2$,$3$, and $4$, yielding
the equally likely Bell state outcomes in Eqs. (\ref{eq7a}, \ref{eq7b}),
project the output modes $5$ and $6$ on the entangled states described in
Eq. (\ref{eq5}). This required joint measurement can be accomplished by the
Bell-state analyzer comprised by Alice's station.

From the Eqs. (\ref{eq1a}, \ref{eq1b}) and (\ref{eq2a}, \ref{eq2b}), it
follows that a measurement through detectors $D_i$ ($i=1$,$2$,$3$,$4$), of
the output state $\left| 0\right\rangle _1\left| 1\right\rangle _2\left|
1\right\rangle _3\left| 0\right\rangle _4$, which requires the incoming Bell
state $\left| \Psi ^{+}\right\rangle _{1234}$, projects the modes $5$ and $6$
entering Bob's station exactly on to the original entangled state injected
through channels $1$ and $2$. After knowing the result of this measurement,
communicated by Alice via the classical channel depicted in Fig. 1(a), Bob
does not need to do anything further to produce a replica of the state $%
\left| \Omega \right\rangle _{12}$. On the other hand, a joint measurement
of the Bell state $\left| \Psi ^{-}\right\rangle _{1234}$ is achieved by
measuring the output state $\left| 1\right\rangle _1\left| 0\right\rangle
_2\left| 1\right\rangle _3\left| 0\right\rangle _4$, leaving the input modes
at Bob's laboratory in the entanglement ${\cal C}_1\left| 0\right\rangle
_5\left| 1\right\rangle _6-{\cal C}_2\left| 1\right\rangle _5\left|
0\right\rangle _6$. In this case, an appropriate unitary transformation has
to be performed by Bob in order to convert this entanglement into a replica
of the original state $\left| \Omega \right\rangle _{12}$. Such a unitary
transformation is accomplished through the application of the operator $%
\widehat{\sigma }_z$, which refers to the Pauli matrix in the basis $\left\{
\left| 0,1\right\rangle _{5,6},\left| 1,0\right\rangle _{5,6}\right\} $.

Regarding the remaining Bell-type measurements, the result $\left| \Phi
^{+}\right\rangle _{1234}$ ($\left| \Phi ^{-}\right\rangle _{1234}$) is
achieved by measuring the output state $\left| 1\right\rangle _{1}\left|
0\right\rangle _{2}\left| 0\right\rangle _{3}\left| 1\right\rangle _{4}$ ($%
\left| 0\right\rangle _{1}\left| 1\right\rangle _{2}\left| 0\right\rangle
_{3}\left| 1\right\rangle _{4}$), leaving the input modes at Bob's
laboratory in the entanglement ${\cal C}_{2}\left| 0\right\rangle _{5}\left|
1\right\rangle _{6}+{\cal C}_{1}\left| 1\right\rangle _{5}\left|
0\right\rangle _{6}$ (${\cal C}_{2}\left| 0\right\rangle _{5}\left|
1\right\rangle _{6}-{\cal C}_{1}\left| 1\right\rangle _{5}\left|
0\right\rangle _{6}$). Here, the required unitary transformation corresponds
to apply the Pauli matrix $\widehat{\sigma }_{x}$ ($\widehat{\sigma }_{y}$)
in the basis $\left\{ \left| 0,1\right\rangle _{5,6},\left| 1,0\right\rangle
_{5,6}\right\} $. The unitary operations $\widehat{\sigma }_{k}$ ($k=z,x,y$%
),\ can be implemented through the quantum resources in Bob's station,
sketched in Fig. 1(d), via appropriate choices of the phase shifts $\phi
_{1} $, $\phi _{2}$, and $\phi _{3}$ introduced by the phase plates. The
outcomes of the photodetections in Alice's station associated with the four
Bell states in Eqs. (\ref{eq7a}, \ref{eq7b}) are summarized in Table 1,
along with the values of the phase shifts required to achieve the rotations $%
\widehat{\sigma }_{k}$ ($k=z,x,y$) in Bob's station. As in the preparation
of the quantum channel, the interaction parameter in the Kerr medium in
Alice's station has been adjusted so that $\chi \tau =\pi $ \cite
{Lukin,Petrosyan}. We finally note, as far as the Bell-type measurements are
concerned, that in a quantum circuit Bob's rotations must be automatically
implemented after the information on the photodetections, received through
classical bits.

\vspace{2.5cm}

\begin{quote}
{\bf Table 1}. In this table we summarize the Bell states, in the first
column, projected by the four possible photodetections arranged in the
second column. The $180^{{{}^{o}}}$ rotations around the $z$, $x$, and $y$
axes, required to convert the resulting entangled modes $5$ and $6$ into a
replica of the original state $\left| \Omega \right\rangle _{12}$, are given
in the third column. The Pauli matrix $\widehat{\sigma }_{k}$ ($k=z,x,y$)
are expressed in the basis $\left\{ \left| 0,1\right\rangle _{5,6},\left|
1,0\right\rangle _{5,6}\right\} $. Finally, in the fourth column we show the
phase shifts $\phi _{1}$, $\phi _{2}$, and $\phi _{3}$ introduced by the
phase plates so as to achieve the required rotations.

\begin{tabular}{cccccccccc}
\hline\hline
BellState &  &  & Photodetection &  &  & Rotations &  &  & Phase Shifts \\ 
\hline
$\left| \Psi ^{+}\right\rangle _{1234}$ & : &  & $\left| 0\right\rangle
_{1}\left| 1\right\rangle _{2}\left| 1\right\rangle _{3}\left|
0\right\rangle _{4}$ &  &  & ${\bf 1}$ &  &  & --- \\ 
$\left| \Psi ^{-}\right\rangle _{1234}$ & : &  & $\left| 1\right\rangle
_{1}\left| 0\right\rangle _{2}\left| 1\right\rangle _{3}\left|
0\right\rangle _{4}$ &  &  & $\widehat{\sigma }_{z}$ &  &  & $\phi _{1}=\phi
_{2}=\phi _{3}=\pi $ \\ 
$\left| \Phi ^{+}\right\rangle _{1234}$ & : &  & $\left| 1\right\rangle
_{1}\left| 0\right\rangle _{2}\left| 0\right\rangle _{3}\left|
1\right\rangle _{4}$ &  &  & $\widehat{\sigma }_{x}$ &  &  & $\phi _{1}=0$, $%
\ \phi _{2}=\phi _{3}=3\pi /2$ \\ 
$\left| \Phi ^{-}\right\rangle _{1234}$ & : &  & $\left| 0\right\rangle
_{1}\left| 1\right\rangle _{2}\left| 0\right\rangle _{3}\left|
1\right\rangle _{4}$ &  &  & $\widehat{\sigma }_{y}$ &  &  & $\phi _{1}=\phi
_{3}=0$, $\phi _{2}=\pi $ \\ \hline\hline
\end{tabular}
\end{quote}

Therefore, we have demonstrated that the proposed scheme for teleporting an
entangled state allows a $100\%$ probability of success in the ideal case,
where only the four Bell states presented in Eqs. (\ref{eq7a}, \ref{eq7b})
occur, out of the sixteen composing the Bell basis described in Eq. (\ref
{eq6}).

\section{Losses in the Optical Elements}

\subsection{Absorptive Beam Splitters}

When the errors due to photoabsorption in the beam splitters are taking into
account, the relationships between the input and output operators $\widehat{%
\alpha },\widehat{\beta }$ (see Fig. 2(a)),{\it \ }described in Eqs. (\ref
{eq1a}, \ref{eq1b}) are generalized to \cite{Jeffers}

\end{mathletters}
\begin{mathletters}
\begin{eqnarray}
\widehat{\alpha }_{out} &=&T\widehat{\alpha }_{in}+R\widehat{\beta }_{in}+%
\widehat{{\cal L}}_\alpha ,  \label{1a} \\
\widehat{\beta }_{out} &=&T\widehat{\beta }_{in}+R\widehat{\alpha }_{in}+%
\widehat{{\cal L}}_\beta ,  \label{1b}
\end{eqnarray}
in order to account for the Langevin noise operator $\widehat{{\cal L}}$
associated with fluctuating currents within the medium composing the beam
splitters. The transmission and reflection coefficients for an absorptive $%
BS $ generalizes those introduced in Eqs. (\ref{eq1a}, \ref{eq1b}) as $T=%
\sqrt{\kappa }t$, $R=\sqrt{\kappa }r$, such that $\left| T\right| ^2+\left|
R\right| ^2=\kappa $, a constant indicating the probability of
nonabsorption, a kind of quality factor for a $BS$. The input fields and the
noise sources are required to be independent, so the input operators must
commute with the Langevin operators

\end{mathletters}
\begin{equation}
\left[ \widehat{\alpha }_{in},\widehat{{\cal L}}_\alpha \right] =\left[ 
\widehat{\alpha }_{in},\widehat{{\cal L}}_\beta \right] =\left[ \widehat{%
\alpha }_{in},\widehat{{\cal L}}_\alpha ^{\dagger }\right] =\left[ \widehat{%
\alpha }_{in},\widehat{{\cal L}}_\beta ^{\dagger }\right] =0,  \label{2}
\end{equation}
with similar relations for the operators $\widehat{\beta }$. Imposition of
the bosonic commutation relations on the output mode operators leads to the
requirements on the commutation relations for the Langevin operators

\begin{mathletters}
\begin{eqnarray}
\left[ \widehat{{\cal L}}_\alpha ,\widehat{{\cal L}}_\alpha ^{\dagger
}\right] &=&\left[ \widehat{{\cal L}}_\beta ,\widehat{{\cal L}}_\beta
^{\dagger }\right] =\Gamma ,  \label{3a} \\
\left[ \widehat{{\cal L}}_\alpha ,\widehat{{\cal L}}_\beta ^{\dagger
}\right] &=&\left[ \widehat{{\cal L}}_\beta ,\widehat{{\cal L}}_\alpha
^{\dagger }\right] =-\Delta ,  \label{3b}
\end{eqnarray}
where $\Gamma =1-\kappa $ is the damping constant and $\Delta
=TR^{*}+RT^{*}=0$. We note that $\Delta $ can assume nonzero values only for
an asymmetric $BS$. At optical frequencies the state of the environment can
be very well approximated by the vacuum state even at room temperature, so
that 
\end{mathletters}
\begin{equation}
\widehat{{\cal L}}_\alpha \left| 0\right\rangle =\widehat{{\cal L}}_\beta
\left| 0\right\rangle =\widehat{\alpha }_{in}\left| 0\right\rangle =\widehat{%
\beta }_{in}\left| 0\right\rangle =0,  \label{4}
\end{equation}
and, from the input-output relations (\ref{1a} and \ref{1b}), it also
follows that

\begin{equation}
\widehat{\alpha }_{out}\left| 0\right\rangle =\widehat{\beta }_{out}\left|
0\right\rangle =0.  \label{5}
\end{equation}
We note that in the above relations $\left| 0\right\rangle $ represents the
vacuum state for modes $\alpha $, $\beta $ and their respective
environments. Finally, the quantum averages of the Langevin operators
vanish, 
\begin{equation}
\left\langle \widehat{{\cal L}}_\alpha \right\rangle =\left\langle \widehat{%
{\cal L}}_\beta \right\rangle =\left\langle \widehat{{\cal L}}_\alpha
^{\dagger }\right\rangle =\left\langle \widehat{{\cal L}}_\beta ^{\dagger
}\right\rangle =0,  \label{6}
\end{equation}
and the ground-state expectation values for the products of pairs of noise
operators are

\begin{mathletters}
\begin{eqnarray}
\left\langle \widehat{{\cal L}}_\alpha \widehat{{\cal L}}_\alpha ^{\dagger
}\right\rangle &=&\left\langle \widehat{{\cal L}}_\beta \widehat{{\cal L}}%
_\beta ^{\dagger }\right\rangle =\Gamma ,  \label{7a} \\
\left\langle \widehat{{\cal L}}_\alpha \widehat{{\cal L}}_\beta ^{\dagger
}\right\rangle &=&\left\langle \widehat{{\cal L}}_\beta \widehat{{\cal L}}%
_\alpha ^{\dagger }\right\rangle =0.  \label{7b}
\end{eqnarray}

As noted in Ref. \cite{Jeffers}, the above relations for the averages of the
Langevin operators may also be derived from a canonical one-dimensional
theory applied to a dielectric slab.

Next, it is readily shown that, similarly to the relations (\ref{1a}) and (%
\ref{1b}), the transformations relating the output to the input operators,
preserving the above-mentioned properties for the Langevin operators, are

\end{mathletters}
\begin{mathletters}
\begin{eqnarray}
\widehat{\alpha }_{in} &=&T^{*}\widehat{\alpha }_{out}+R^{*}\widehat{\beta }%
_{out}+\widehat{{\cal L}}_\alpha ,  \label{8a} \\
\widehat{\beta }_{in} &=&T^{*}\widehat{\beta }_{out}+R^{*}\widehat{\alpha }%
_{out}+\widehat{{\cal L}}_\beta ,  \label{8b}
\end{eqnarray}
where the bosonic commutation relations are satisfied by the input mode
operators. From Eqs. (\ref{8a},\ref{8b}), the output state arising from the
injection of a photon through mode $\alpha $ of a Beam splitter is given by 
\end{mathletters}
\begin{equation}
\left| 1,0\right\rangle _{\alpha \beta }^{in}=\widehat{\alpha }%
_{in}^{\dagger }\left| 0,0\right\rangle _{\alpha \beta }^{in}=\left[ T\left|
1,0\right\rangle _{\alpha \beta }^{out}+R\left| 0,1\right\rangle _{\alpha
\beta }^{out}+\widehat{{\cal L}}_\alpha ^{\dagger }\left| 0,0\right\rangle
_{\alpha \beta }^{out}\right] \left| {\bf 0}\right\rangle _{{\bf E}},
\label{9}
\end{equation}
where $\left| {\bf 0}\right\rangle _{{\bf E}}=\prod_k\left| 0\right\rangle
_k=$ $\left| \left\{ 0_k\right\} \right\rangle $ stands for the state of the
environment composed of a huge number of vacuum-field states $\left|
0\right\rangle _k$.

\subsection{Absorptive Kerr Medium}

To deal with photoabsorption in a Kerr medium we again take advantage of the
Langevin operators. Similarly to the procedure adopted above, to introduce
photoabsorption into a $BS$, the coupling between the input and output
operators $\widehat{\alpha }$, $\widehat{\beta }$ (see Fig. 2(b)),{\it \ }%
described in an ideal Kerr medium by Eqs. (\ref{eq2a}, \ref{eq2b}), is
generalized in a lossy Kerr medium to 
\begin{mathletters}
\begin{eqnarray}
\widehat{\alpha }_{out} &=&\sqrt{\eta }\exp \left( i\chi \tau \widehat{\beta 
}_{in}^{\dagger }\widehat{\beta }_{in}\right) \widehat{\alpha }_{in}+%
\widehat{{\frak L}}_{\alpha }{\rm {,}}  \label{E1a} \\
\widehat{\beta }_{out} &=&\sqrt{\eta }\exp \left( i\chi \tau \widehat{\alpha 
}_{in}^{\dagger }\widehat{\alpha }_{in}\right) \widehat{\beta }_{in}+%
\widehat{{\frak L}}_{\beta }{\rm {,}}  \label{E1b}
\end{eqnarray}
where $\Lambda =1-\eta $ is the probability of photoabsorption, the damping
constant for the Kerr medium. (We use different characters for the Langevin
operators, to differentiate the particular optical element responsible for
the photoabsorption.)

The algebraic rules satisfied by the input and output operators $\widehat{%
\alpha }$, $\widehat{\beta }$ and the Langevin operators are similar to
those for the $BS$. Assuming the commutation of the Langevin operators with
the input operators $\widehat{\alpha }_{in}$, $\widehat{\beta }_{in}$, and
imposing bosonic commutation rules on the output mode operators $\widehat{%
\alpha }_{out}$, $\widehat{\beta }_{out}$, it follows that

\end{mathletters}
\begin{mathletters}
\begin{eqnarray}
\left[ \widehat{{\frak L}}_\alpha ,\widehat{{\frak L}}_\alpha ^{\dagger
}\right] &=&\left[ \widehat{{\frak L}}_\beta ,\widehat{{\frak L}}_\beta
^{\dagger }\right] =\Lambda ,  \label{E2a} \\
\left[ \widehat{{\frak L}}_\alpha ,\widehat{{\frak L}}_\beta ^{\dagger
}\right] &=&\left[ \widehat{{\frak L}}_\beta ,\widehat{{\frak L}}_\alpha
^{\dagger }\right] =0.  \label{E2b}
\end{eqnarray}
Approximating the state of the environment by the vacuum state, we obtain
relations analogous to those in Eqs.(\ref{4})-(\ref{6}) for the Kerr medium
operators. The ground-state expectation values for the products of pairs of
noise operators are thus 
\end{mathletters}
\begin{mathletters}
\begin{eqnarray}
\left\langle \widehat{{\frak L}}_\alpha \widehat{{\frak L}}_\alpha ^{\dagger
}\right\rangle &=&\left\langle \widehat{{\frak L}}_\beta \widehat{{\frak L}}%
_\beta ^{\dagger }\right\rangle =\Lambda ,  \label{E3a} \\
\left\langle \widehat{{\frak L}}_\alpha \widehat{{\frak L}}_\beta ^{\dagger
}\right\rangle &=&\left\langle \widehat{{\frak L}}_\beta \widehat{{\frak L}}%
_\alpha ^{\dagger }\right\rangle =0.  \label{E3b}
\end{eqnarray}
The transformations relating the output to the input operators, preserving
all the above-mentioned properties, are

\end{mathletters}
\begin{mathletters}
\begin{eqnarray}
\widehat{\alpha }_{in} &=&\sqrt{\eta }\exp \left( -i\chi \tau \widehat{\beta 
}_{out}^{\dagger }\widehat{\beta }_{out}\right) \widehat{\alpha }_{out}+%
\widehat{{\frak L}}_\alpha {\rm {,}}  \label{E4a} \\
\widehat{\beta }_{in} &=&\sqrt{\eta }\exp \left( -i\chi \tau \widehat{\alpha 
}_{out}^{\dagger }\widehat{\alpha }_{out}\right) \widehat{\beta }_{out}+%
\widehat{{\frak L}}_\beta {.}  \label{E4b}
\end{eqnarray}

When computing the output state arising from the injection of two photons
through modes $\alpha $ and $\beta $ of a Kerr medium, we obtain form Eqs. (%
\ref{E4a}, \ref{E4b}) 
\end{mathletters}
\begin{eqnarray}
\left| 1,1\right\rangle _{\alpha \beta }^{in} &=&\widehat{\alpha }%
_{in}^{\dagger }\widehat{\beta }_{in}^{\dagger }\left| 0,0\right\rangle
_{\alpha \beta }^{in}=\left[ \eta e^{-i\chi \tau }\left| 1,1\right\rangle
_{\alpha \beta }^{out}+\sqrt{\eta }\left| 0,1\right\rangle _{\alpha \beta
}^{out}\widehat{{\frak L}}_{\alpha }^{\dagger }\right.  \nonumber \\
&&\left. +\sqrt{\eta }\left| 1,0\right\rangle _{\alpha \beta }^{out}\widehat{%
{\frak L}}_{\beta }^{\dagger }+\left| 0,0\right\rangle _{\alpha \beta }^{out}%
\widehat{{\frak L}}_{\alpha }^{\dagger }\widehat{{\frak L}}_{\beta
}^{\dagger }\right] \left| 0\right\rangle _{{\bf E}},  \label{E5}
\end{eqnarray}
where $\left| 0\right\rangle _{{\bf E}}$ denotes the initial state of the
environment. In this equation, note that when considering the ideal case, $%
\eta =1$, we correctly obtain $\left| 1,1\right\rangle _{\alpha \beta
}^{in}=e^{-i\chi \tau }\left| 1,1\right\rangle _{\alpha \beta }^{out}$. In
an absorptive Kerr medium, we find that the expected value for the states $%
\left| 0,1\right\rangle _{\alpha \beta }^{out}$ or $\left| 1,0\right\rangle
_{\alpha \beta }^{out}$ is correctly given by $\eta \left( 1-\eta \right) $,
while the probability for the absorption of both photons is $\left( 1-\eta
\right) ^{2}$. We stress that a detailed treatment of dissipation in a Kerr
medium should consider the time intervals, between zero\ and $\tau $ (the
interaction time within an ideal Kerr medium), for the absorption of photons
in modes $\alpha $\ and $\beta $; as a consequence, additional phase factors
would appear in Eq. (\ref{E5}). However, the above treatment is sufficient
for providing a good estimative about the fidelity of the proposed
teleportation process.

\subsection{Efficiency of the Detectors}

Introducing output operators to account for the detection of a given input
field $\alpha $ reaching the detectors, we have

\begin{equation}
\widehat{\alpha }_{out}=\sqrt{\varepsilon }\widehat{\alpha }_{in}+\widehat{%
{\sf L}}_\alpha ,  \label{E9}
\end{equation}
where $\varepsilon $ stands for the efficiency of the detector. Obviously,
differently from the case of the $BS$ and Kerr medium, the detectors do not
couple different modes. The Langevin operators $\widehat{{\sf L}}_\alpha $,
besides satisfying all the properties of those introduced above, obey the
commutation relations

\begin{mathletters}
\begin{eqnarray}
\left[ \widehat{{\sf L}}_\alpha ,\widehat{{\sf L}}_\alpha ^{\dagger }\right]
&=&1-\varepsilon ,  \label{E10a} \\
\left[ \widehat{{\sf L}}_\alpha ,\widehat{{\sf L}}_\beta ^{\dagger }\right]
&=&0,  \label{E10b}
\end{eqnarray}
and the ground-state expectation values for the products of pairs are

\end{mathletters}
\begin{mathletters}
\begin{eqnarray}
\left\langle \widehat{{\sf L}}_\alpha \widehat{{\sf L}}_\alpha ^{\dagger
}\right\rangle &=&1-\varepsilon ,  \label{E11a} \\
\left\langle \widehat{{\sf L}}_\alpha \widehat{{\sf L}}_\beta ^{\dagger
}\right\rangle &=&0.  \label{E11b}
\end{eqnarray}

\subsection{Absorptive phase plates}

The photoabsorption in the phase plates is modeled by analogy with the above
treatment for the efficiency of the detectors. Introducing an output
operator for the $\alpha $-mode of a phase plate with damping constant $%
1-\varkappa $, it follows that

\end{mathletters}
\begin{equation}
\widehat{\alpha }_{out}=\sqrt{\varkappa }\widehat{\alpha }_{in}+\widehat{%
{\Bbb L}}_\alpha ,  \label{E110}
\end{equation}
with the Langevin operator $\widehat{{\Bbb L}}_\alpha $ obeying similar
relations to those in Eqs. (\ref{E10a}, \ref{E10b}) and (\ref{E11a}, \ref
{E11b}).

\subsection{General Relations for the Errors due to bean splitters and
Detectors}

For the sake of generality, we next introduce relations accounting for both
sources of error: photoabsorption in the $BS$ (Eqs. (\ref{1a}) and (\ref{1b}%
)) and the efficiency of detectors (Eq. (\ref{E9})). One can prove that in
this formulation the output operators $\widehat{\alpha },\widehat{\beta }$,
which describe the output fields from $BS_{5{\rm {\ }}}$and $BS_{6{\rm {\ }}%
} $reaching the detectors, are

\begin{mathletters}
\begin{eqnarray}
\widehat{\alpha }_{out} &=&{\bf T}\widehat{\alpha }_{in}+{\bf R}\widehat{%
\beta }_{in}+\widehat{{\bf L}}_\alpha ,  \label{E12a} \\
\widehat{\beta }_{out} &=&{\bf T}\widehat{\beta }_{in}+{\bf R}\widehat{%
\alpha }_{in}+\widehat{{\bf L}}_\beta ,  \label{E12b}
\end{eqnarray}
where ${\bf T=}\sqrt{\varepsilon }T$, ${\bf R=}\sqrt{\varepsilon }R$, and $%
\widehat{{\bf L}}_\alpha =\widehat{{\cal L}}_\alpha +\widehat{{\frak L}}%
_\alpha $. In fact, combining all the above-mentioned properties of the
operators in relations (\ref{E12a}) and (\ref{E12b}), we obtain

\end{mathletters}
\begin{mathletters}
\begin{eqnarray}
\left[ \widehat{{\bf L}}_\alpha ,\widehat{{\bf L}}_\alpha ^{\dagger }\right]
&=&\left[ \widehat{{\bf L}}_\beta ,\widehat{{\bf L}}_\beta ^{\dagger
}\right] =\varepsilon \Gamma +\left( 1-\varepsilon \right) ,  \label{E13a} \\
\left[ \widehat{{\bf L}}_\alpha ,\widehat{{\bf L}}_\beta ^{\dagger }\right]
&=&\left[ \widehat{{\bf L}}_\beta ,\widehat{{\bf L}}_\alpha ^{\dagger
}\right] =0.  \label{E13b}
\end{eqnarray}

When substituting $\eta =1$ in (\ref{E13a}) and (\ref{E13b}), we recover the
relations (\ref{3a}) and (\ref{3b}), while for $\Gamma =0$, we recover the
relations (\ref{E10a}) and (\ref{E10b}), respectively.

\section{Teleportation of an Entangled State: Noise Effects}

In calculating the fidelity of the teleportation process, we assume that the
state to be teleported, $\left| \Omega \right\rangle _{12}$, is prepared
ideally, with fidelity equal to unity. However, by taking into account the
damping constants of the three $BS$ and the Kerr medium involved in
preparing the quantum channel, as depicted in Fig. 1(b), we obtain the
nonideal entanglement

\end{mathletters}
\begin{eqnarray}
\left| \widetilde{\Upsilon }\right\rangle _{3456} &=&\left\{ \xi
^{3/2}\left[ \left( 1-\eta ^{1/2}\right) \left| 1\right\rangle _3\left|
0\right\rangle _4+i\eta ^{1/2}\left( 1+\eta ^{1/2}\right) \left|
0\right\rangle _3\left| 1\right\rangle _4\right] \left| 1\right\rangle
_5\left| 0\right\rangle _6\right.  \nonumber \\
&&+\xi ^{3/2}\left[ i\left( 1+\eta ^{1/2}\right) \left| 1\right\rangle
_3\left| 0\right\rangle _4-\eta ^{1/2}\left( 1-\eta ^{1/2}\right) \left|
0\right\rangle _3\left| 1\right\rangle _4\right] \left| 0\right\rangle
_5\left| 1\right\rangle _6  \nonumber \\
&&+\xi ^{1/2}\left[ \left( \widehat{{\cal L}}_5^{(2)\dagger }+i\xi
^{1/2}\eta ^{1/2}\widehat{{\cal L}}_5^{(3)\dagger }+i\xi ^{1/2}\widehat{%
{\frak L}}_5^{\dagger }+\xi ^{1/2}\widehat{{\cal L}}_6^{(3)\dagger }\right)
\left| 1\right\rangle _3\left| 0\right\rangle _4\right.  \nonumber \\
&&+\left. i\eta ^{1/2}\left( \widehat{{\cal L}}_5^{(2)\dagger }-i\xi
^{1/2}\eta ^{1/2}\widehat{{\cal L}}_5^{(3)\dagger }+i\xi ^{1/2}\widehat{%
{\frak L}}_5^{\dagger }+\xi ^{1/2}\widehat{{\cal L}}_6^{(3)\dagger }\right)
\left| 0\right\rangle _3\left| 1\right\rangle _4\right] \left|
0\right\rangle _5\left| 0\right\rangle _6  \nonumber \\
&&\left. +\left| \vartheta \right\rangle _{3456}\right\} \left| {\bf 0}%
\right\rangle _{{\bf E}}{\rm {,}}  \label{E14}
\end{eqnarray}
where $\xi =\kappa /2$ and the superscript $\ell $ of the Langevin operators 
$\widehat{{\cal L}}^{(\ell )}$ refers to the $\ell $th beam splitter, and
thus the $\ell $th environment where the photon has been absorbed, so that $%
\left[ \widehat{{\cal L}}_\alpha ^{(\ell )},\widehat{{\cal L}}_\alpha
^{(\ell ^{\prime })\dagger }\right] =\delta _{\ell \ell ^{\prime }}\Gamma $.
Note that $\left| {\bf 0}\right\rangle _{{\bf E}}$ is the product of all the
environments, referred to each $BS$ and $KM_1$, i.e., $\left| {\bf 0}%
\right\rangle _{{\bf E}}=\left| \left\{ 0_k\right\} \right\rangle
_{BS_1}\left| \left\{ 0_k\right\} \right\rangle _{BS_2}\left| \left\{
0_k\right\} \right\rangle _{BS_3}\left| \left\{ 0_k\right\} \right\rangle
_{KM_1}$. There is no need to introduce a superscript to label the Langevin
operator accounting for the absorptive Kerr media, since they can be
differentiated only through their respective modes. The ket $\left|
\vartheta \right\rangle _{3456}$, which involves only zero-photon states in
modes $3$ and $4$, is expanded in Appendix A.

The use of four photodetectors to achieve the Bell measurement enables us to
probe the occurrence of photoabsorptions in the optical elements placed
between the input and output channels $1$ to $4$. Since one photon is
injected through mode $1$ or $2$, composing the state to be teleported, and
another is injected through mode $3$ or $4$, composing part of the quantum
channel, Alice communicates to Bob only the successful events, where the
former photon is detected through $D_{1}$ or $D_{2}$ and the latter through $%
D_{3}$ or $D_{4}$. As far as the photon injected through modes $3$ or $4$ is
concerned, these successful events include only the states explicitly shown
in superposition (\ref{E14}), since the remaining ket $\left| \vartheta
\right\rangle _{3456}$ contains only zero-photon states in modes $3$ and $4$%
. Evidently, when one of the photons (from the channel couples $1$-$2$ or $3$%
-$4$) or both of them have been absorbed or scattered by the optical
elements, that event is disregarded and the teleportation process must be
restarted. This strategy, in association with the usually small damping
constants characterizing beam splitters and Kerr media, results in a
high-fidelity teleportation process. Such fidelity is not affected by the
efficiency of the photodetectors, which plays a role only in the probability
of the Bell state measurements: in the real situation, each of the four
measurement outcomes occurs with probability smaller then $1/4$, as
discussed below. We note that the above strategy does not rule out the error
coming from dark counts in the detectors, which can be treated according to
Ref. \cite{Ozdemir}. However, the error due to dark counts will be effective
only in two cases: {\it i)} when a photon from the channel couples $1$-$2$
or $3$-$4$ is absorbed and a dark count occurs simultaneously to the
detection of the remaining nonabsorbed photon, and {\it ii)} when both
photons from the channel couples $1$-$2$ and $3$-$4$ are absorbed and two
dark counts occur simultaneously in the detector couples $D_{1}-D_{2}$ and $%
D_{3}-D_{4}$. All other possibilities for dark counts are excluded by the
above strategy, resulting in an rather small error due to false signals.

Starting with the fidelity of the non-ideal quantum channel in Eq. (\ref{E14}%
), i.e., the fidelity of the reduced density operator ${\rm {Tr}}_{{\bf E}%
}\left( \left| \widetilde{\Upsilon }\right\rangle \left\langle \widetilde{%
\Upsilon }\right| \right) $relative to the ideal quantum channel $\left|
\Upsilon \right\rangle $, we obtain the expression

\begin{equation}
{\cal F}=\left\langle \Upsilon \left| {\rm {Tr}}_{{\bf E}}\left( \left| 
\widetilde{\Upsilon }\right\rangle \left\langle \widetilde{\Upsilon }\right|
\right) \right| \Upsilon \right\rangle =\frac{\xi ^{3}}{2}\left[ 1+4\eta
^{1/2}(1+\eta )+\eta (6+\eta )\right] .  \label{E15}
\end{equation}
As expected, when disregarding the losses in bean splitters $1$,$2$ and $3$ (%
$\kappa =2\xi =1$) and in the $KM_{1}$ ($\eta =1$) we find ${\cal F}=1$.

We now consider, for simplicity, the realistic situation where the photon
states $\left| 0\right\rangle _1\left| 1\right\rangle _2\left|
1\right\rangle _3\left| 0\right\rangle _4$ are detected, corresponding to
Alice's measurement, in the ideal case, of the Bell state $\left| \Psi
^{+}\right\rangle _{1234}$. As mentioned above, the measurement of the
output state $\left| 0\right\rangle _1\left| 1\right\rangle _2\left|
1\right\rangle _3\left| 0\right\rangle _4$ projects the modes $5$ and $6$
entering Bob's station exactly on to the original entangled state $\left|
\Omega \right\rangle _{12}$, so that Bob need not do anything further to
obtain a replica of this state in the ideal case. Assuming realistic
non-ideal optical elements, the output modes $5$ and $6$ in Alice's station,
before the photodetection, read

\begin{equation}
\left| \psi \right\rangle _{56}={\cal N}\left[ {\bf a}\left| 0\right\rangle
_5\left| 1\right\rangle _6+{\bf b}\left| 1\right\rangle _5\left|
0\right\rangle _6+\left( {\bf c}\widehat{{\cal L}}_5^{(2)\dagger }+{\bf d}%
\widehat{{\cal L}}_5^{(3)\dagger }+{\bf e}\widehat{{\frak L}}_5^{\dagger }+%
{\bf f}\widehat{{\cal L}}_6^{(3)\dagger }\right) \left| 0\right\rangle
_5\left| 0\right\rangle _6\right] \left| {\bf 0}\right\rangle _{{\bf E}}{\rm 
{,}}  \label{E16}
\end{equation}
where the normalization constant is 
\begin{equation}
{\cal N}=\left[ \left| {\bf a}\right| ^2+\left| {\bf b}\right| ^2+\left(
1-2\xi \right) \left( \left| {\bf c}\right| ^2+\left| {\bf d}\right|
^2+\left| {\bf f}\right| ^2\right) +\left| {\bf e}\right| ^2\left( 1-\eta
\right) \right] ^{-1/2}{\rm {.}}  \label{E17}
\end{equation}
Therefore, the probability of detecting the state $\left| 0\right\rangle
_1\left| 1\right\rangle _2\left| 1\right\rangle _3\left| 0\right\rangle _4$,
given by ${\sf P}_{0110}={\cal N}^{-2}\varepsilon ^2$ (as can be obtained
from the evolution of the of the state $\left| \Omega \right\rangle
_{12}\otimes \left| \widetilde{\Upsilon }\right\rangle _{3456}$ through
Alice's station), depends on the efficiency of the photodetectors and turns
out to be $1/4$ when assuming ideal $BS$'s ($\kappa =2\xi =1$), $KM$'s ($%
\eta =1$), and photodetectors ($\varepsilon =1$). In fact, for the
coefficients in Appendix B, we obtain ${\bf a}=$ $-{\cal C}_1/2$ and ${\bf b}%
=$ $-{\cal C}_2/2$, so that ${\cal N}=\left[ \left| {\bf a}\right| ^2+\left| 
{\bf b}\right| ^2\right] ^{-1/2}=2$.

The coefficients ${\bf a}$ to ${\bf f}$ are displayed in Appendix B. From
Eq. (\ref{E16}), the reduced density operator can be obtained: 
\begin{eqnarray}
\widehat{\rho }_{56} &=&{\rm {Tr}}_{{\bf E}}\left| \psi \right\rangle
_{56}\left\langle \psi \right| ={\cal N}\left\{ \left| {\bf a}\right|
^2\left| 0,1\right\rangle _{56}\left\langle 0,1\right| +{\bf ab}^{*}\left|
0,1\right\rangle _{56}\left\langle 1,0\right| \right.  \nonumber \\
&&+{\bf a}^{*}{\bf b}\left| 1,0\right\rangle _{56}\left\langle 0,1\right|
+\left| {\bf b}\right| ^2\left| 1,0\right\rangle _{56}\left\langle 1,0\right|
\nonumber \\
&&+\left. \left[ \left( 1-2\xi \right) \left( \left| {\bf c}\right|
^2+\left| {\bf d}\right| ^2+\left| {\bf f}\right| ^2\right) +\left| {\bf e}%
\right| ^2\left( 1-\eta \right) \right] \left| 0,0\right\rangle
_{56}\left\langle 0,0\right| \right\} {\rm {.}}  \label{E18}
\end{eqnarray}
Finally, the fidelity of the teleportation process, the overlap between the
ideal state $\left| \Omega \right\rangle _{56}={\cal C}_1\left|
0\right\rangle _5\left| 1\right\rangle _6+{\cal C}_2\left| 1\right\rangle
_5\left| 0\right\rangle _6$ and the non-ideal teleported state in Eq. (\ref
{E16}), is given by

\begin{eqnarray}
{\sf F} &=&\left. _{56}\left\langle \Omega \right| \right. \widehat{\rho }%
_{56}\left| \Omega \right\rangle _{56}  \nonumber \\
&=&{\cal N}^2\left( \left| {\bf a}\right| ^2\left| {\cal C}_1\right| ^2+{\bf %
ab}^{*}{\cal C}_1^{*}{\cal C}_2+{\bf a}^{*}{\bf b}{\cal C}_1{\cal C}%
_2^{*}+\left| {\bf b}\right| ^2\left| {\cal C}_2\right| ^2\right) .
\label{E19}
\end{eqnarray}
It follows immediately that, for the ideal case where $\kappa =\eta
=\varepsilon =2\xi =1$, so that ${\bf a}=$ $-{\cal C}_1/2$, ${\bf b}=$ $-%
{\cal C}_2/2$, and ${\cal N}=2$, we obtain ${\sf F}=\left( \left| {\cal C}%
_1\right| ^2+\left| {\cal C}_2\right| ^2\right) ^2=1$. Remember that the
function {\sf F} given here is the fidelity of the teleported state
associated with Alice's measurement of the Bell state $\left| \Psi
^{+}\right\rangle _{1234}$. We note that the normalization factor ${\cal N}$%
, and so the fidelity ${\sf F}$, would be considerably smaller when
considering events where no photon is detected in either of the channel
couples $1$-$2$ or $3$-$4$.

Evidently, different expressions for the fidelity of the teleported state
follow from different results of the Bell measurement. In fact, Bob's
intervention on the state entering his station, after Alice's measurement,
must be specific according to Table 1, and the more the optical elements
used to accomplish the rotation required to convert the teleported state
into a replica of the original entanglement $|\Omega \rangle _{12}$, the
more the errors introduced in the teleported state. In this connection, we
expect an increasing fidelity for the teleported entanglement, when
detecting the Bell states $\left| \Psi ^{-}\right\rangle _{1234}$, $\left|
\Phi ^{+}\right\rangle _{1234}$, $\left| \Phi ^{-}\right\rangle _{1234}$, $%
\left| \Psi ^{+}\right\rangle _{1234}$, in that order.

An important point to be stressed is that the strategy for probing the
absorption of photons between the inputs and outputs of channels $1$ to $4$
is insufficient to eliminate the introduction of errors by the optical
elements between these channels. In fact, even when both photons injected
through modes $1$ or $2$ and $3$ or $4$ are not absorbed, the errors
introduced by the optical elements will invalidate Table 1, in the sense
that the measurement of the state $\left| 0\right\rangle _{1}\left|
1\right\rangle _{2}\left| 1\right\rangle _{3}\left| 0\right\rangle _{4}$,
for example, turns to be associated not only with Bell state $\left| \Psi
^{+}\right\rangle _{1234}$ (as in the ideal case), but also to the three
other Bell states, although with smaller probabilities. In Appendix C we
present the evolution of the Bell states $\left| \Psi ^{\pm }\right\rangle
_{1234}$ (appearing in the expansion of the product $\left| \Omega
\right\rangle _{12}\otimes \left| \widetilde{\Upsilon }\right\rangle _{3456}$%
) through Alice's station, showing the mixing of all the possibilities of
photodetections in Table 1.

\section{Comments and Conclusion}

In this paper we have presented a scheme for the teleportation of an
EPR-type entanglement of zero- and one-photon running-wave states. Besides
employing linear optical elements, such as beam splitters and phase plates,
our teleportation apparatus also incorporates Kerr media to allow a $100\%$
probability of success in the ideal case. A scheme for teleporting similar
entanglements of running-wave states, employing only linear optical
elements, was recently suggested \cite{Kim,Lee}, but the theoretical
probability of success was only $50\%$. Following the lines suggested in 
\cite{Kim}, the experimental teleportation of an entangled qubit was
recently carried out \cite{Martini}. In this connection, we point out that
the Kerr nonlinearity required in our proposal can be achieved with
presently available technology, given the recently reported ultraslow light
propagation in cold atomic media \cite{Lukin,Petrosyan,Hau}. Lukin and
Imamoglu \cite{Lukin} have demonstrated that a conditional phase shift of
the order of $\pi $ could be achieved if both light pulses, propagating with
slow but equal group velocity in a coherently-prepared atomic gas, were
submitted to electromagnetically induced transparency. It is worth
mentioning the increasing research interest in achieving giant crossed-Kerr
nonlinearity through ultraslow light propagation in a cold gas of atoms \cite
{Lukin,Hau}. Such activity encourage theoretical propositions such as the
present one and the recently reported scheme for complete quantum
teleportation of the polarization state of a photon, employing Kerr
nonlinearity, which also requires a conditional phase shift $\pi $ \cite
{Tombesi}. In Ref. \cite{Petrosyan} it is suggested a scheme that allows
equal, slow group velocities of the interacting photons and therefore a
cross-phase shift of $\pi $, which cannot be achieved using the scheme in
Ref. \cite{Lukin} owing to the group velocity mismatch of the two photons.

We have also provided a phenomenological approach to compute the influence
of damping in Kerr media, reasoning by analogy with the treatment of lossy
beam splitters given in \cite{Jeffers}. The efficiency of the photodetectors
has also been introduced, as well as the influence of damping in phase
plates, by making use of the relations previously developed in \cite{Celso}.
Therefore, a comprehensive treatment of errors in the domain of running-wave
physics, for linear and nonlinear optical elements, has been presented which
allows the fidelity of the teleportation process to be computed (in the
particular situation where the result of Alice's Bell-type measurement
prevents Bob from the necessity to perform any appropriate unitary
transformation on the received state). The strategy employed to carry out
the Bell measurement is able to probe the occurrence of photoabsorptions in
the optical elements. This strategy, combined with the usually small damping
constants of the beam splitters and Kerr media and the high efficiency of
the photodetectors, results in a high-fidelity teleportation process (as
analyzed below).

To estimate the fidelities of the prepared quantum channel (Eq. (\ref{E15}))
and the teleported state (Eq. (\ref{E19})), we note that the efficiency of
single-photon detectors is about $70\%$, yielding $\varepsilon \simeq 0.7$,
while the damping constant for a $BS$ is rather small, less than $2\%$ in $%
BK7$ crystals, given $\kappa \simeq 0.98$. Considering light pulses of tiny
energy crossing the Kerr medium, the methods proposed in Refs. \cite
{Lukin,Petrosyan} to achieve a nonlinear phase shift of the order of $\pi $
permits the interaction to be maintained for a very long time without
dissipation. Hence, for the Kerr medium composed of a cold gas of atoms, the
damping constant is also very small, and for the purpose of this example we
assume the same value as for the $BS$, so that $\eta \simeq 0.98$. As
stressed above, we have computed in Eq. (\ref{E19}) the fidelity of the
teleported state corresponding to Alice's measurement, in the ideal case, of
the Bell state $\left| \Psi ^{+}\right\rangle _{1234}$. This choice was
motivated only to simplify the calculation of the fidelity since the three
other Bell states require appropriated rotations accomplished in {\it Bob's
station }through phase plates and additional Beam Splitters. With these
considerations we obtain for the fidelity of the non-ideal quantum channel $%
{\cal F}=0.92$. The fidelity {\sf F} of the teleported state, given by Eq. (%
\ref{E19}), is plotted in Fig. $3$ as a function of parameters $\gamma $ and 
$\lambda $ defining the coefficients ${\cal C}_{1}=\cos $($\gamma $) and $%
{\cal C}_{2}=\sin $($\gamma $)$%
\mathop{\rm e}%
\nolimits^{i\lambda }$. As is evident from Fig. $3$, the fidelity of the
teleported state for $\gamma =0$($\pi $), when $|\Omega \rangle _{12}=\left|
0\right\rangle _{1}\left| 1\right\rangle _{2}$($-\left| 0\right\rangle
_{1}\left| 1\right\rangle _{2}$), is smaller than that for $\gamma =\pi /2$,
when $|\Omega \rangle _{12}=\left| 1\right\rangle _{1}\left| 0\right\rangle
_{2}$. In fact, when the photon goes through mode $2$ ($\gamma =0$,$\pi $),
it crosses $KM_{2}$, introducing errors into the process which do not occur
when the photon travels through mode $1$($\gamma =\pi /2$). On the other
hand, for a superposition state, i.e., $\gamma \neq 0$,$\pi $, we note that
the phase factor $%
\mathop{\rm e}%
\nolimits^{i\lambda }$\ plays an important role in the fidelity. Even for $%
\gamma =\pi /4$, when the photon in the state to be teleported has equal
probabilities of travelling through modes $1$ or $2$, the interference
process occurring in $BS_{6}$, depending on the phase factor $%
\mathop{\rm e}%
\nolimits^{i\lambda }$, ensures different values for the fidelity {\sf F}.
The reason is that the function {\sf F} gives the fidelity of the teleported
state associated with Alice's measurement of the Bell state $\left| \Psi
^{+}\right\rangle _{1234}$ and the phase factor $%
\mathop{\rm e}%
\nolimits^{i\lambda }$ leads to different probabilities for the output
photodetection $\left| 0\right\rangle _{1}\left| 1\right\rangle _{2}\left|
1\right\rangle _{3}\left| 0\right\rangle _{4}$ associated with $\left| \Psi
^{+}\right\rangle _{1234}$. As shown in Appendix C, the errors introduced by
the optical elements mix together the Bell states associated with a given
output photodetection and the probabilities of measuring each of the Bell
states will depend on $\lambda $.

From the values fixed above for the efficiency of the detectors and the
probability of nonabsorption of the beam splitters and the Kerr media, we
can also estimate the probability ${\sf P}_{0110}={\cal N}^{-2}\varepsilon
^2 $ of detecting the output state $\left| 0\right\rangle _1\left|
1\right\rangle _2\left| 1\right\rangle _3\left| 0\right\rangle _4$ in
Alice's station. When considering ${\cal C}_1=$ ${\cal C}_2=1/\sqrt{2}$, the
probability ${\sf P}_{0110}$, which in the ideal case is $0.25$, turns to be
about $0.11$. For the efficiency of the detectors equal to unity, we obtain $%
{\sf P}_{0110}={\cal N}^{-2}\approx 0.22$. Note that the factor $\varepsilon
^2$, which reduces the probability ${\sf P}_{0110}$ from $0.22$ to $0.11$,
follows from the necessity of detecting both photons in modes $2$ and $3$.

In addition to the above-mentioned errors arising from the absorptive beam
splitters, Kerr media, and phase plates, there are other experimental
nonidealities that seem to be important. In fact, the two modes interfering
at the beam splitters are never matched perfectly and the effects of the
mode mismatch can be discussed within the multimode theory \cite{Banaszek}.
Moreover, we have to account for the fact that we do not have perfect
single-photon sources. The commonly cited method of parametric fluorescence
only approximates a single-photon source, and this approximation must be
evaluated for its effect on the fidelity of the teleported state. The
interaction between the two light pulses crossing the Kerr medium is also
subject to errors due to fluctuations of the associated physical parameters.
Finally, it is worth mentioning that the errors arising from fluctuating
parameters could be taken into account through the recently proposed
phenomenological-operator technique \cite{POT}.

{\bf Acknowledgments}

We wish to express thanks for the support from FAPESP (under contracts
\#98/03171-9, \#99/11617-0 and \#00/15084-5) and CNPq (Intituto do
Mil\^{e}nio de Informa\c{c}\~{a}o Qu\^{a}ntica), Brasilian agencies.

{\large Appendix A}

The ket $\left| \vartheta \right\rangle _{3456}$, in Eq. (\ref{E14}),
involving only zero-photon states in modes $3$ and $4$ reads

\begin{eqnarray*}
\left| \vartheta \right\rangle _{3456} &=&\left\{ \xi ^2\left[ \left( 1-\eta
^{1/2}\right) +i\xi -i\xi \eta ^{1/2}\widehat{{\frak L}}_4^{\dagger }\right]
\left| 1\right\rangle _5\left| 0\right\rangle _6\right. \\
&&+i\xi ^2\left[ \left( 1+\eta ^{1/2}\right) +\xi +\xi \eta ^{1/2}\widehat{%
{\frak L}}_4^{\dagger }\right] \left| 0\right\rangle _5\left| 1\right\rangle
_6 \\
&&+\left[ \left( \widehat{{\cal L}}_5^{(2)\dagger }+i\xi \eta ^{1/2}\widehat{%
{\cal L}}_5^{(3)\dagger }+i\xi \widehat{{\frak L}}_5^{\dagger }+\xi \widehat{%
{\cal L}}_6^{(3)\dagger }\right) \right. \\
&&+\left. \left. i\xi \left( \widehat{{\cal L}}_5^{(2)\dagger }+i\xi \eta
^{1/2}\widehat{{\cal L}}_5^{(3)\dagger }+i\xi \widehat{{\frak L}}_5^{\dagger
}+\xi \widehat{{\cal L}}_6^{(3)\dagger }\right) \widehat{{\frak L}}%
_4^{\dagger }\right] \left| 0\right\rangle _5\left| 0\right\rangle
_6\right\} \left| 0\right\rangle _3\left| 0\right\rangle _4\left| {\bf 0}%
\right\rangle _{{\bf E}}{\rm {.}}
\end{eqnarray*}

{\large Appendix B}

The coefficients related to the teleported state in Eq. (\ref{E16}) are
given by

\begin{eqnarray*}
{\bf a} &=&\xi ^3\left[ -\eta ^{1/2}\left( 1+\eta ^{1/2}+3\eta -\eta
^{3/2}\right) {\cal C}_1+i\left( 1-\eta ^{1/2}-\eta +\eta ^{3/2}\right) 
{\cal C}_2\right] {\rm {,}} \\
{\bf b} &=&\xi ^3\left[ i\eta ^{1/2}\left( 1-\eta ^{1/2}-\eta +\eta
^{3/2}\right) {\cal C}_1+\left( 1-3\eta ^{1/2}-\eta -\eta ^{3/2}\right) 
{\cal C}_2\right] {\rm {,}}
\end{eqnarray*}

\begin{eqnarray*}
{\bf c} &=&\xi ^{2}\left[ i\eta ^{1/2}\left( 1+\eta \right) {\cal C}%
_{1}+\left( 1-2\eta ^{1/2}-\eta \right) {\cal C}_{2}\right] {\rm {,}} \\
{\bf d} &=&\xi ^{5/2}\left\{ -\eta \left[ 1+2\eta ^{1/2}-\eta \right] {\cal C%
}_{1}+i\eta ^{1/2}\left( 1+\eta \right) {\cal C}_{2}\right\} {\rm {,}} \\
{\bf e} &=&\xi ^{5/2}\left\{ -\eta ^{1/2}\left[ 1+\frac{1}{2}\eta
^{1/2}-\eta ^{1/2}(1-\eta ^{1/2})+\eta \right] {\cal C}_{1}\right. \\
&&\left. +i\left[ 1-\frac{3}{2}\eta ^{1/2}-\eta ^{1/2}(1+\eta ^{1/2})-\frac{1%
}{2}\eta \right] {\cal C}_{2}\right\} {\rm {,}} \\
{\bf f} &=&\xi ^{5/2}\left[ i\eta ^{1/2}\left( 1-\eta \right) {\cal C}%
_{1}+\left( 1-2\eta ^{1/2}-\eta \right) {\cal C}_{2}\right] .
\end{eqnarray*}

{\large Appendix C}

The evolution of the Bell states $\left| \Psi ^{\pm }\right\rangle _{1234}$
(appearing in the expansion of the product $\left| \Omega \right\rangle
_{12}\otimes \left| \widetilde{\Upsilon }\right\rangle _{3456}$) through
Alice's station (Fig. 1(c)) is given by

\[
\left| \Psi ^{\pm }\right\rangle _{1234}=\frac 1{\sqrt{2}}\left( \left|
0,1\right\rangle _{12}\left| 1,0\right\rangle _{34}\pm \left|
1,0\right\rangle _{12}\left| 0,1\right\rangle _{34}\right) \qquad \qquad
\qquad \qquad 
\]
\begin{eqnarray*}
&&\stackrel{Alice^{\prime }s.station}{\longrightarrow }\frac 1{\sqrt{2}}%
\left\{ \xi ^{3/2}\left[ \left( \eta ^{1/2}+\eta \right) \mp \left( \eta
^{1/2}+1\right) \right] \left| 1,0\right\rangle _{12}\left| 1,0\right\rangle
_{34}\right. \\
&&\qquad \qquad +i\xi ^{3/2}\left[ \left( \eta ^{1/2}+\eta \right) \pm
\left( \eta ^{1/2}+1\right) \right] \left| 0,1\right\rangle _{12}\left|
1,0\right\rangle _{34} \\
&&\qquad \qquad +\xi ^{3/2}\left[ -\left( \eta ^{1/2}-\eta \right) \pm
\left( \eta ^{1/2}-1\right) \right] \left| 0,1\right\rangle _{12}\left|
0,1\right\rangle _{34} \\
&&\qquad \qquad +i\xi ^{3/2}\left[ \left( \eta ^{1/2}-\eta \right) \pm
\left( \eta ^{1/2}-1\right) \right] \left| 1,0\right\rangle _{12}\left|
0,1\right\rangle _{34} \\
&&\qquad \qquad \left. +\left| \vartheta _{\Psi ^{\pm }}\right\rangle
_{1234}\right\} {\rm {,}}
\end{eqnarray*}
where the ket $\left| \vartheta _{\Psi ^{\pm }}\right\rangle _{1234}$,
presenting zero-photon states in at least one of the channel couples $1$-$2$
or $3$-$4$, reads 
\begin{eqnarray*}
\left| \vartheta _{\Psi ^{\pm }}\right\rangle _{1234} &=&\left\{ \xi \left[
\left( \eta ^{1/2}+\eta \right) \widehat{{\cal L}}_2^{(6)\dagger }\left|
1,0\right\rangle _{34}+i\left( \eta ^{1/2}-\eta \right) \widehat{{\cal L}}%
_2^{(6)\dagger }\left| 0,1\right\rangle _{34}\right. \right. \\
&&+\left( \widehat{{\frak L}}_2^{\dagger }-\eta ^{1/2}\widehat{{\frak L}}%
_2^{\dagger }\right) \left| 1,0\right\rangle _{34}+i\left( \widehat{{\frak L}%
}_2^{\dagger }+\eta ^{1/2}\widehat{{\frak L}}_2^{\dagger }\right) \left|
0,1\right\rangle _{34} \\
&&\pm \left. \left( \eta ^{1/2}-1\right) \widehat{{\cal L}}_1^{(6)\dagger
}\left| 0,1\right\rangle _{34}\pm i\left( \eta ^{1/2}+1\right) \widehat{%
{\cal L}}_1^{(6)\dagger }\left| 1,0\right\rangle _{34}\right] \left|
0,0\right\rangle _{12} \\
&&+\left[ \left( \xi \eta \right) ^{1/2}\left( \widehat{{\cal L}}%
_3^{(4)\dagger }-i\left( \xi \eta \right) ^{1/2}\widehat{{\cal L}}%
_3^{(5)\dagger }+i\xi ^{1/2}\widehat{{\frak L}}_3^{\dagger }+\xi ^{1/2}%
\widehat{{\cal L}}_4^{(5)\dagger }\right) \left( \left| 1,0\right\rangle
_{12}+i\left| 0,1\right\rangle _{12}\right) \right. \\
&&\pm \left. \xi ^{1/2}\left( \left( \xi \eta \right) ^{1/2}\widehat{{\cal L}%
}_3^{(5)\dagger }+\xi ^{1/2}\widehat{{\frak L}}_3^{\dagger }+\widehat{{\cal L%
}}_4^{(4)\dagger }+i\xi ^{1/2}\widehat{{\cal L}}_4^{(5)\dagger }\right)
\left( \left| 0,1\right\rangle _{12}+i\left| 1,0\right\rangle _{12}\right)
\right] \left| 0,0\right\rangle _{34} \\
&&+\left[ \eta ^{1/2}\left( \widehat{{\cal L}}_3^{(4)\dagger }-i\left( \xi
\eta \right) ^{1/2}\widehat{{\cal L}}_3^{(5)\dagger }+i\xi ^{1/2}\widehat{%
{\frak L}}_3^{\dagger }+\xi ^{1/2}\widehat{{\cal L}}_4^{(5)\dagger }\right) 
\widehat{{\cal L}}_2^{(6)\dagger }\right. \\
&&+\left( \widehat{{\cal L}}_3^{(4)\dagger }+i\left( \xi \eta \right) ^{1/2}%
\widehat{{\cal L}}_3^{(5)\dagger }+i\xi ^{1/2}\widehat{{\frak L}}_3^{\dagger
}+\xi ^{1/2}\widehat{{\cal L}}_4^{(5)\dagger }\right) \widehat{{\frak L}}%
_2^{\dagger } \\
&&\left. \left. \pm \left( \left( \xi \eta \right) ^{1/2}\widehat{{\cal L}}%
_3^{(5)\dagger }+\xi ^{1/2}\widehat{{\frak L}}_3^{\dagger }+\widehat{{\cal L}%
}_4^{(4)\dagger }+i\xi ^{1/2}\widehat{{\cal L}}_4^{(5)\dagger }\right) 
\widehat{{\cal L}}_1^{(6)\dagger }\right] \left| 0,0\right\rangle
_{12}\left| 0,0\right\rangle _{34}\right\} \left| {\bf 0}\right\rangle _{%
{\bf E}}{\rm {.}}
\end{eqnarray*}

For the realistic values $\kappa =\eta =0.98$ we have

\begin{eqnarray*}
&&\left| \Psi ^{+}\right\rangle _{1234}\stackrel{Alice^{\prime }s.station}{%
\longrightarrow }i\left( 0.9604\right) \left| 0,1\right\rangle _{12}\left|
1,0\right\rangle _{34}-\left( 0.0049\right) \left| 0,1\right\rangle
_{12}\left| 0,1\right\rangle _{34} \\
&&\qquad \qquad \qquad \qquad -\left( 0.0049\right) \left| 1,0\right\rangle
_{12}\left| 1,0\right\rangle _{34}+\left| \vartheta _{\Psi
^{+}}\right\rangle _{1234}{\rm {,}}
\end{eqnarray*}

\begin{eqnarray*}
&&\left| \Psi ^{-}\right\rangle _{1234}\stackrel{Alice^{\prime }s.station}{%
\longrightarrow }\left( 0.9604\right) \left| 1,0\right\rangle _{12}\left|
1,0\right\rangle _{34}-i\left( 0.0049\right) \left| 0,1\right\rangle
_{12}\left| 1,0\right\rangle _{34} \\
&&\qquad \qquad \qquad \qquad +\left( 0.0049\right) \left| 1,0\right\rangle
_{12}\left| 0,1\right\rangle _{34}+\left| \vartheta _{\Psi
^{-}}\right\rangle _{1234}{\rm {,}}
\end{eqnarray*}
showing the breakdown of the one-to-one relation between the photodetections
and Bell states presented in Table 1 for the ideal case. A similar analysis
applies to Bell states $\left| \Phi ^{\pm }\right\rangle _{1234}$.

\noindent{\bf Figure Caption}

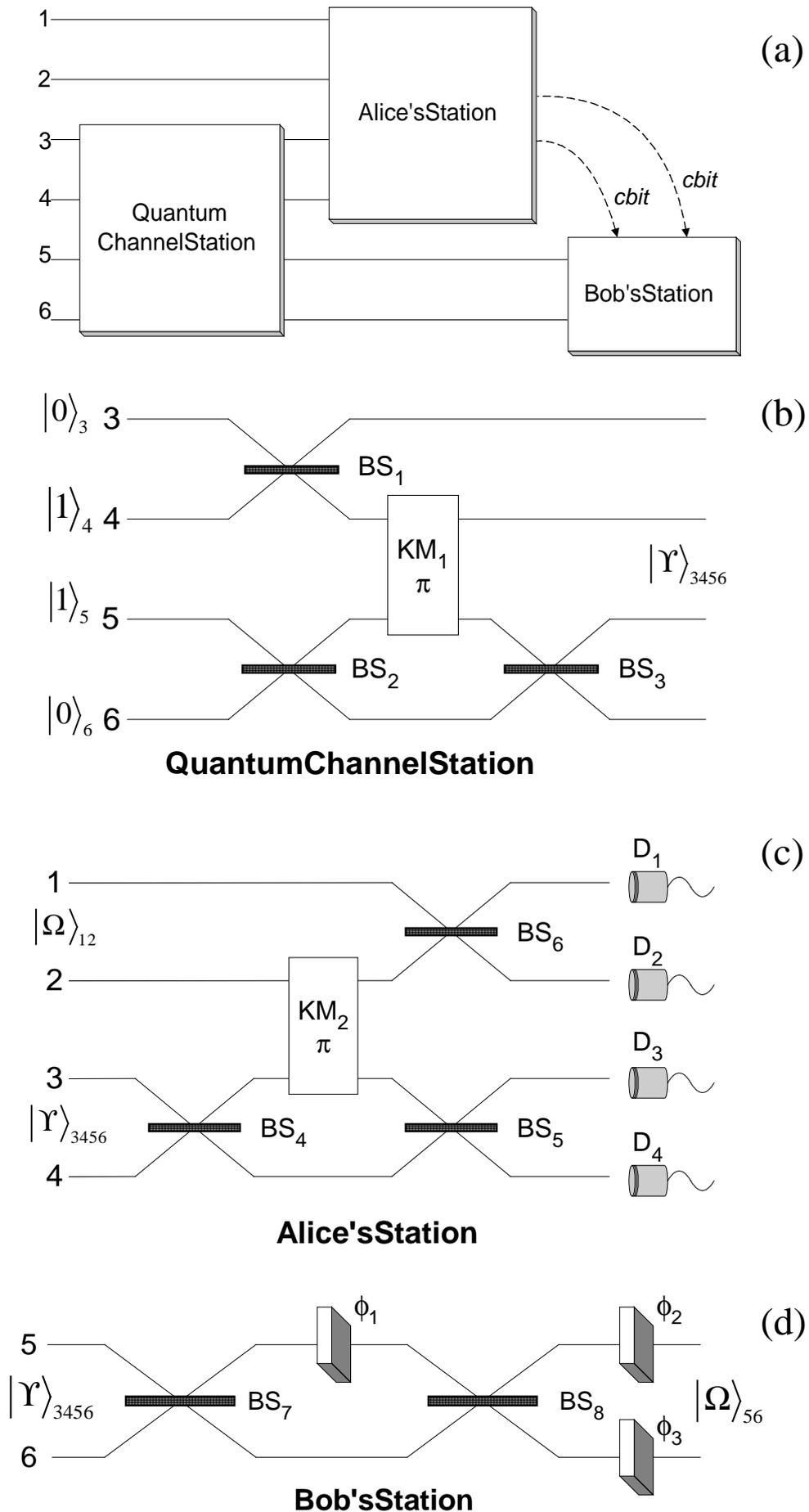
FIG. 1. (a) Sketch of the experimental setup for quantum teleportation of
entanglements of zero- and one-photon running-wave states. The apparatus is
composed of a station to prepare the quantum channel, Alice's station for
carrying out the Bell-type measurement, and Bob's station to accomplish all
the rotations required to convert the teleported state into a replica of the
original entangled state. The dashed lines indicate the classical channel
composed of two classical bits (cbit). (b) Sketch of the station used to
engineer the quantum channel, an entanglement of modes $3$ to $6$ which is
shared by Alice and Bob. The output modes $3$ and $4$ enter Alice's station
to be entangled with modes $1$ and $2$ ( encapsulating the state to be
teleported). The output modes $5$ and $6$, which receive the teleported
state after Alice's Bell-type measurement, enter Bob's station for the
accomplishment, if needed, of the rotation required to convert the
teleported state into a replica of the original entanglement $|\Omega
\rangle _{12}$. The {\it quantum channel station} consists of a beam
splitter ($BS_{1}$), a MZ interferometer, composed of a pair of beam
splitters ($BS_{2}$, $BS_{3}$), and a Kerr medium ($KM_{1}$). (c) Sketch of 
{\it Alice's station} which consists of the same ingredients as the station
used to engineer the quantum channel plus photodetectors $D_{i}$ ($i=1$,$2$,$%
3$,$4$). The lack of a single photodetection event in couples $D_{1}$-$D_{2}$
or $D_{3}$-$D_{4}$ would indicate that a photon, either from the state to be
teleported or the quantum channel, has been absorbed (between the input and
output channels $1$ to $4$). The events where a single photon or both of
them have been absorbed or scattered by the optical elements are disregarded
and the teleportation process must be restarted. This strategy contributes
to the higher fidelity of the present teleportation scheme, since we do not
compute the terms in Eq. (\ref{E14}) for the nonideal quantum channel
(associated to ket $\left| \vartheta \right\rangle _{3456}$), which involve
only zero-photon states in modes $3$ and $4$. (d) Sketch of {\it Bob's
station,} where a MZ interferometer ($BS_{7}$, $BS_{8}$) and three phase
plates are employed to accomplish all the rotations required to convert the
teleported state into a replica of the original entangled state $|\Omega
\rangle _{12}$, allowing a $100\%$ probability of success in the ideal case.

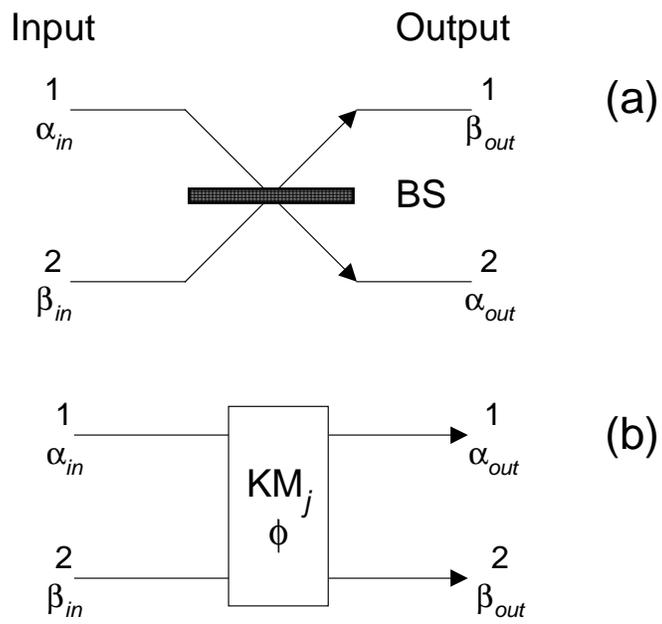
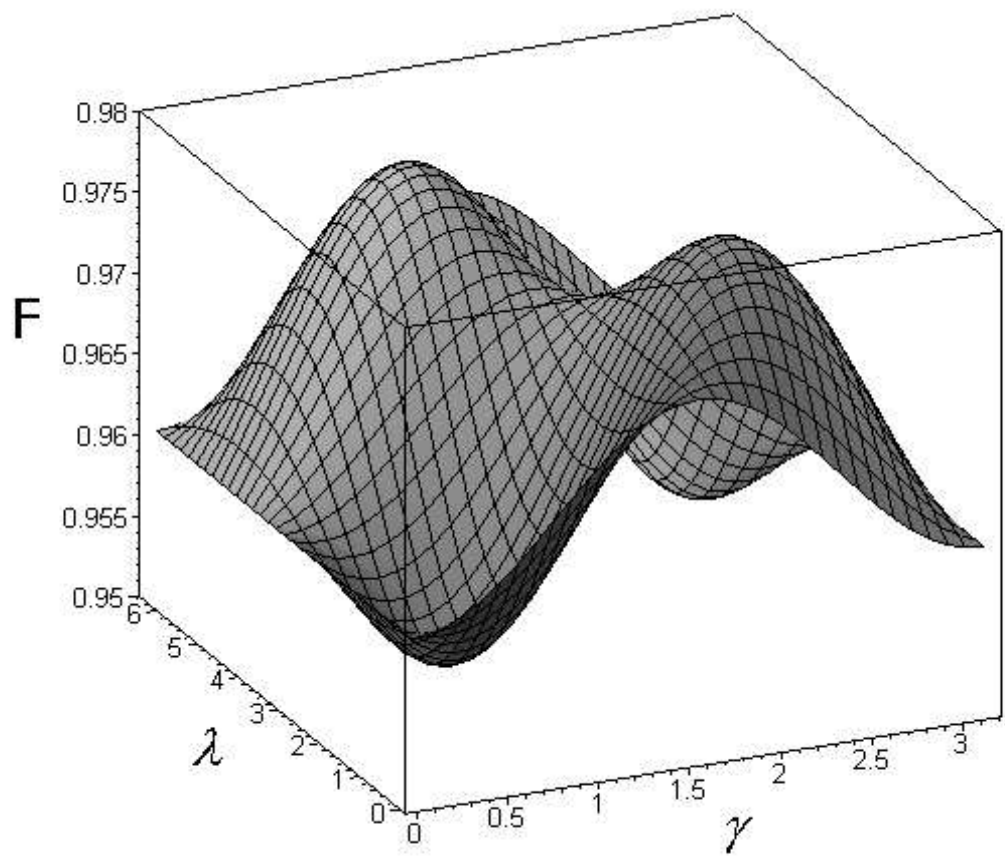
FIG. 2. Schematic representation of input and output modes for (a) the Beam
Splitters and (b) the Kerr Media.

FIG. 3. Fidelity {\sf F} of the teleported state (associated with Alice's
measurement of the Bell state $\left| \Psi ^{+}\right\rangle _{1234}$),
expressed in Eq. (\ref{E19}), as a function of parameters $\gamma $ and $%
\lambda $ defining the coefficients ${\cal C}_{1}=\cos $($\gamma $) and $%
{\cal C}_{2}=\sin $($\gamma $)$%
\mathop{\rm e}%
\nolimits^{i\lambda }$.

\end{document}